\documentstyle[epsfig]{aprim10}
\input{epsf}

\newif\ifAMStwofonts

\ifoldfss
  \ifCUPmtlplainloaded \else
    \NewTextAlphabet{textbfit} {cmbxti10} {}
    \NewTextAlphabet{textbfss} {cmssbx10} {}
    \NewMathAlphabet{mathbfit} {cmbxti10} {} 
    \NewMathAlphabet{mathbfss} {cmssbx10} {} 
  \fi
  \ifAMStwofonts
    \ifCUPmtlplainloaded \else
      \NewSymbolFont{upmath} {eurm10}
      \NewSymbolFont{AMSa} {msam10}
      \NewMathSymbol{\upi}     {0}{upmath}{19}
      \NewMathSymbol{\umu}     {0}{upmath}{16}
      \NewMathSymbol{\upartial}{0}{upmath}{40}
      \NewMathSymbol{\leqslant}{3}{AMSa}{36}
      \NewMathSymbol{\geqslant}{3}{AMSa}{3E}

    \fi
  \fi
\fi 

\ifnfssone
  \newmathalphabet{\mathit}
  \addtoversion{normal}{\mathit}{cmr}{m}{it}
  \addtoversion{bold}{\mathit}{cmr}{bx}{it}
  \newmathalphabet{\mathbfit} 
  \addtoversion{normal}{\mathbfit}{cmr}{bx}{it}
  \addtoversion{bold}{\mathbfit}{cmr}{bx}{it}
  \newmathalphabet{\mathbfss} 
  \addtoversion{normal}{\mathbfss}{cmss}{bx}{n}
  \addtoversion{bold}{\mathbfss}{cmss}{bx}{n}
  \ifAMStwofonts
    \ifCUPmtlplainloaded \else
      \UseAMStwoboldmath
      \makeatletter
      \new@mathgroup\upmath@group
      \define@mathgroup\mv@normal\upmath@group{eur}{m}{n}
      \define@mathgroup\mv@bold\upmath@group{eur}{b}{n}
      \edef\UPM{\hexnumber\upmath@group}
      \new@mathgroup\amsa@group
      \define@mathgroup\mv@normal\amsa@group{msa}{m}{n}
      \define@mathgroup\mv@bold\amsa@group{msa}{m}{n}
      \edef\AMSa{\hexnumber\amsa@group}
      \makeatother
      \mathchardef\upi="0\UPM19
      \mathchardef\umu="0\UPM16
      \mathchardef\upartial="0\UPM40
      \mathchardef\leqslant="3\AMSa36
      \mathchardef\geqslant="3\AMSa3E
    \fi
  \fi
\fi 

\ifnfsstwo
  \DeclareMathAlphabet{\mathbfit}{OT1}{cmr}{bx}{it}
  \SetMathAlphabet\mathbfit{bold}{OT1}{cmr}{bx}{it}
  \DeclareMathAlphabet{\mathbfss}{OT1}{cmss}{bx}{n}
  \SetMathAlphabet\mathbfss{bold}{OT1}{cmss}{bx}{n}
  \ifAMStwofonts
    \ifCUPmtlplainloaded \else
      \DeclareSymbolFont{UPM}{U}{eur}{m}{n}
      \SetSymbolFont{UPM}{bold}{U}{eur}{b}{n}
      \DeclareSymbolFont{AMSa}{U}{msa}{m}{n}
      \DeclareMathSymbol{\upi}{0}{UPM}{"19}
      \DeclareMathSymbol{\umu}{0}{UPM}{"16}
      \DeclareMathSymbol{\upartial}{0}{UPM}{"40}
      \DeclareMathSymbol{\leqslant}{3}{AMSa}{"36}
      \DeclareMathSymbol{\geqslant}{3}{AMSa}{"3E}
    \fi
  \fi
\fi 

\ifCUPmtlplainloaded \else
  \ifAMStwofonts \else 
    \def\upi{\pi}
    \def\umu{\mu}
    \def\upartial{\partial}
  \fi
\fi

\title[UT's Dichroic-Mirror Camera (DMC)]{Science with UT's Dichroic-Mirror Camera (DMC), 15-band simultaneous imager}

\author[Kuncarayakti et al.]
       {Hanindyo Kuncarayakti$^1$, Mamoru Doi$^2$, Hakim L. Malasan$^1$,  
       \newauthor Junji Hayano$^2$, Hiroyuki Utsunomiya$^2$, Yutaka Ihara$^2$, Kouichi Tokita$^2$,  
       \newauthor Naohiro Takanashi$^{2,4}$, Shigeyuki Sako$^2$, Sadanori Okamura$^3$, Tomoki Morokuma$^4$, 
       \newauthor Hisanori Furusawa$^4$, Yutaka Komiyama$^4$, Masafumi Yagi$^4$, Norio Okada$^4$,  
       \newauthor Hidehiko Nakaya$^4$, Akira Arai$^5$, Makoto Uemura$^5$, Koji S. Kawabata$^5$,   
       \newauthor Takuya Yamashita$^5$, Takashi Ohsugi$^5$, Masanao Abe$^6$, Sunao Hasegawa$^6$ \\
        $^1$Bosscha Observatory \& Department of Astronomy, Institut Teknologi Bandung, Bandung, Indonesia\\
        $^2$Institute of Astronomy, Graduate School of Science, The University of Tokyo, Japan \\
        $^3$Department of Astronomy, Graduate School of Science, The University of Tokyo, Japan \\
				$^4$National Astronomical Observatory of Japan\\
				$^5$Hiroshima University, Japan\\
				$^6$Japan Aerospace Exploration Agency }
\date{}

\pagerange{\pageref{firstpage}--\pageref{lastpage}}
\pubyear{2008}

\begin{document}

\maketitle

\label{firstpage}

\begin{abstract}
We introduce the Dichroic-Mirror Camera (DMC), an instrument developed at the University of Tokyo which is capable of performing simultaneous imaging in fifteen bands. The main feature of the DMC is the dichroic mirrors, which split incoming light into red and blue components. Combination of dichroic mirrors split light from the telescope focus into fifteen intermediate-width bands across 390 -- 950 nm. The fifteen bands of DMC provide measurements of the object's spectral energy distribution (SED) at fifteen wavelength points. During May -- June 2007 and March 2008, observing run of the DMC was carried out at Higashi-Hiroshima Observatory, Japan. We observed several objects i.e. planets, asteroids, standard stars \& star clusters, planetary nebulae, galaxies, and supernovae. We describe several early scientific results from the DMC.
\end{abstract}


\section{Introduction}

Photometry and spectroscopy have been used extensively in studying the nature of celestial objects of interest. The use of CCD detectors also greatly improves the performance of data acquisition system. While generally having the advantage of producing images of celestial objects -- thus providing spatial information --, CCD photometry lacks the relatively high wavelength resolution compared to spectroscopy. On the other hand, spectroscopy could provide astronomers with detailed information on spectra of celestial objects yet no or very little spatial information could be obtained from the normal spectrographs. 

In order to obtain both spatial and spectral information of an object, we propose the use of a camera using dichroic mirrors to split light into several bands recorded by imaging detectors. By splitting the light from the object we hope to obtain adequate spectral resolution while simultaneously attain images of the object in several bands. We have developed a first version of this kind of instrument at the Institute of Astronomy, the University of Tokyo, Japan. In this presentation we would like to briefly describe the design of our Dichroic-Mirror Camera (DMC), observations using DMC, and first results.

\section{Instrument design}

The main optical component of the DMC is the dichroic mirrors, which could split incoming light into blue (shorter wavelength) and red (longer wavelength) components. Combination of fourteen dichroic mirrors and several other optical components in DMC such as flat mirrors and collimator lenses splits incoming light into fifteen photometric bands across 390 -- 950 nm wavelength region. Design of the instrument has been described in Doi et al. \shortcite{doi98}. Since we could not optimize the configuration of the CCD detectors, the optical design of DMC is quite complicated.

\begin{figure} 
 \centerline{{\epsfxsize=3 cm\epsffile{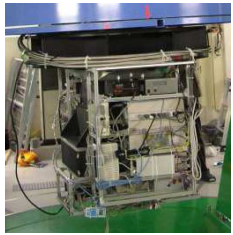}}}
 \caption[]{The DMC, attached to the Cassegrain focus of Kanata telescope.}
 \label{dmc}
\end{figure}

We used a mozaic CCD camera which was previously used as the detector for the Kiso 1.05m Schmidt camera \cite{sekiguchi92}. Fifteen TI TC-215 CCDs of 1000 $\times$ 1018 pixels were positioned at the ends of the light paths to record images in respective wavelengths. Each pixel measured 12 $\times$ 12 microns. The fifteen bands of DMC provide measurements of the object's spectral energy distribution (SED) at fifteen wavelength points simultaneously. This DMC performance can be regarded as some kind of low-resolution multi-object imaging-spectroscopy. Although dichroic mirrors could obtain $\sim80\%$ efficiency, the average throughput of the instrument in each band is about 10\%. This is caused by the old, thick CCD we used which have a relatively low quantum efficiency peaking in $\sim50\%$ at 600 -- 700 nm. This performance of 10\% efficiency of 15-band imager is considered equivalent to 150\% efficiency of a single band imager. By making use of the mozaic CCD, DMC also has the advantage of 15-band simultaneous imaging capability. We have replaced the original CCD electronics with MESSIA5 and MFRONT, developed at NAOJ. Typical laboratory value of CCD readout noise is 15 e$^{-}$ r.m.s. We used a low-cost, portable Stirling-cycle cooler MA-SCU08 by Shinei to cool the CCD to under $-80^{\circ}$ C. Typical pressure inside the CCD dewar is in the order of $10^{-5}$ Torr. The instrument, including shutter and a guider camera, is controlled by using an onboard Linux PC, which in turn controlled by another one via LAN. The whole instrument weighs about 150 kg and measured about 1 m on its sides. Figure \ref{dmc} shows the appearance of the DMC.

\section{Observations}
During May - June 2007 the first observing run of DMC was carried out at Higashi-Hiroshima Observatory, Japan \cite{doi08}. Attached to the Cassegrain focus of 1.5m (f/12.2) Kanata telescope, each CCD of the DMC provides field of view of 4.41', with pixel scale of 0.27". We observed several objects i.e. planets, asteroids, stars \& star clusters, planetary nebulae, galaxies, and supernovae. The second observation run was carried out in March 2008 using the same telescope. In this second run we made several improvements i.e. better optical alignment and lower CCD temperature. This observing session is also a commissioning run for the guider camera, which enabled exposures up to $\sim$30 minutes. In this run we also tried to employ grism mode in DMC to obtain spectra in fifteen bands. Using grism, we managed to reach spectral resolution $R \sim 100$. Unfortunately, one dichroic mirror was accidentally detached from the instrument, thus observation could only be performed in fourteen bands. We obtained S/N = 10 in 5 minutes exposure for 18-mag object (AB magnitude, at 600 nm). Based on standard star measurements, we estimate photometric accuracy of our instrument to be about 2\%.

\begin{figure} 
 \centerline{{\epsfxsize=4 cm\epsffile{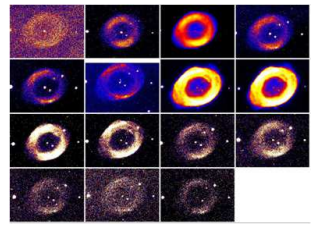}}}
 \caption[]{One set of planetary nebula M57 images. The bluest band is the most upper-left image and wavelength increases rightward, along rows.}
 \label{m57}
\end{figure}

In Figure \ref{m57} we show an example of a set of images. These images of planetary nebula M57 immediately show the wavelength dependance. Third image from left, uppermost row is in the region of [O III] line, around 500 nm. The third and fourth images from the left, second row, are in H$\alpha$ region.

\section{Results}
\begin{figure} 
 \centerline{{\epsfxsize=5.5 cm\epsffile{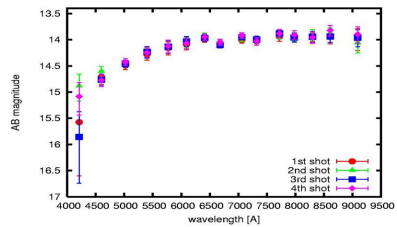}}}
 \caption[]{SED of asteroid 747 Winchester.}
 \label{winc}
\end{figure}

As DMC has been producing raw data images, some of them have been reduced to scientific results. One example is a study of asteroid 747 Winchester \cite{utsunomiya08}. Winchester is an asteroid with rapid rotation, period of 9.4 hour. Our photometry with DMC in 15 bands enabled us to determine the SED of the object. Based on our SED measurements, Winchester was classified as a type C asteroid (Figure \ref{winc}). DMC has been proved to be powerful in SED study like the above example, where high-resolution spectroscopy would be less efficient considering the observing time. Typical exposure time for Winchester was 2 to 3 minutes. Another example is a study of star clusters by Kuncarayakti \shortcite{hk08}. We observed globular cluster M13 and open cluster Ruprecht 8 using DMC to derive their physical parameters. M13 is a well-studied cluster, which could provide an assessment on how good the DMC photometry is compared to other methods of study. In the other hand, Ru 8 is a sparsely populated galactic cluster with no previous study found in the literatures. 

\begin{figure} 
 \centerline{{\epsfxsize=4.5 cm\epsffile{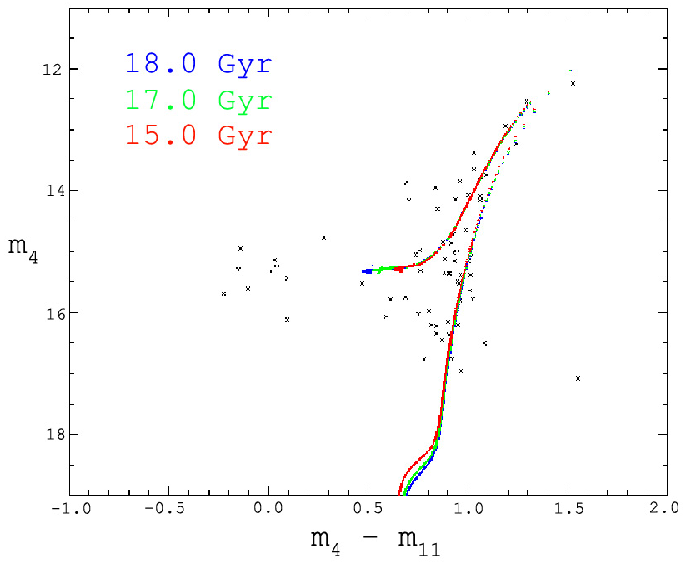}}
 {\epsfxsize=4.5 cm\epsffile{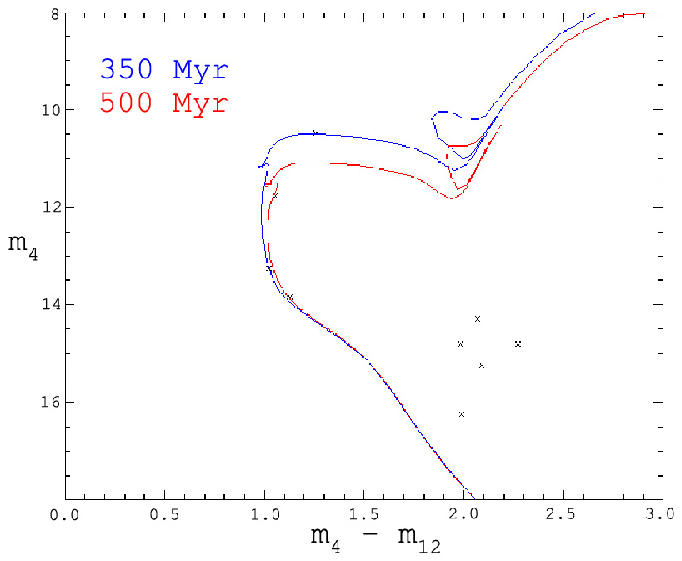}}}
 \caption[]{CMD of M13 (left) and Ru 8 (right).}
 \label{cmd}
\end{figure}

\begin{figure} 
 \centerline{{\epsfxsize=5.3 cm\epsffile{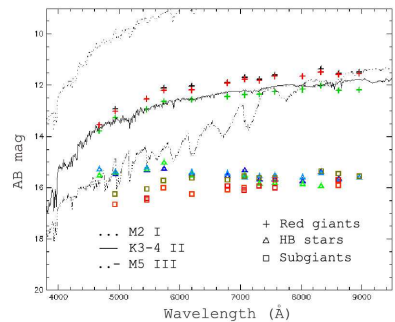}}}
 \caption[]{SED of several M13 stars.}
 \label{sed13}
\end{figure}

For M13 we observed the cluster in four different regions around the cluster center -- this is to avoid the central part where crowding is most severe. Fourteen-band photometry resulted in SED of cluster stars as well as photometric Hertzprung-Russel Diagram, or color-magnitude diagram (CMD) of the cluster. Figure \ref{cmd} shows a CMD of M13 and Figure \ref{sed13} shows SED of several brightest stars. By performing isochrone fitting techniques and checked further with comparing observed and reference SED, we obtained physical parameters of the cluster $T$ = 18.0 Gyr, [Fe/H] = -1.67, $(m_4-M_4)$ = 14.9, $E(m_4-m_{11})$ = 0.0, $d$ = 9.5 kpc. We used low-metallicity isochrones from Pisa Evolutionary Library \cite{castellani03} for this work. Reference SEDs were those of Pickles \shortcite{pickles98} scaled to match the absolute magnitudes defined by Drilling \& Landolt \shortcite{drilling00}. The derived result, however, does not agree perfectly with values from references. The age of M13 is about 12 Gyr according to Rey et al. \shortcite{rey01}, and Catalog of Parameters for Milky Way Globular Clusters \cite{harris96} gives [Fe/H] = -1.54 and $d$ = 7.7 kpc. The rather large scatter in the cluster CMD is suspected to be the reason why our result disagrees with the reference. The photometry was also did not go deep enough to reach the main-sequence turn-off point, where age determination is more definitive. Instrument noise and inaccurate photometric calibrations are some of the causes of the scatter in CMD. Despite this, we could see in Figure \ref{sed13} that our photometry managed to disentangle between different types of stars according to their position in the cluster HR Diagram, as attested by the shape of their SED. To study a star cluster, a method of SED fitting could be a powerful alternative to isochrone fitting as shown by e.g. Wu et al. \shortcite{wu04}.

For Ru 8 we observed the central region of the cluster. The resulting CMD is very poorly populated. Applying the same isochrone technique as M13, we found $T$ = 350-500 Myr, [Fe/H] = -0.76, $(m_4-M_4)$ = 11.54, $E(m_4-m_{12})$ = 1.05, $d$ = 800 pc. We used Padova isochrones \cite{bertelli94} for this purpose. According to the fit, the brightest stars of this cluster should be spectral type A stars. However, the SED of its stars does not seem to agree with classification based on CMD (Figure 4). We used the same reference SED set as in M13. This result indicates that the cluster still deserves further study to reveal it doubtful nature.

\begin{figure} 
 \centerline{{\epsfxsize=5.3 cm\epsffile{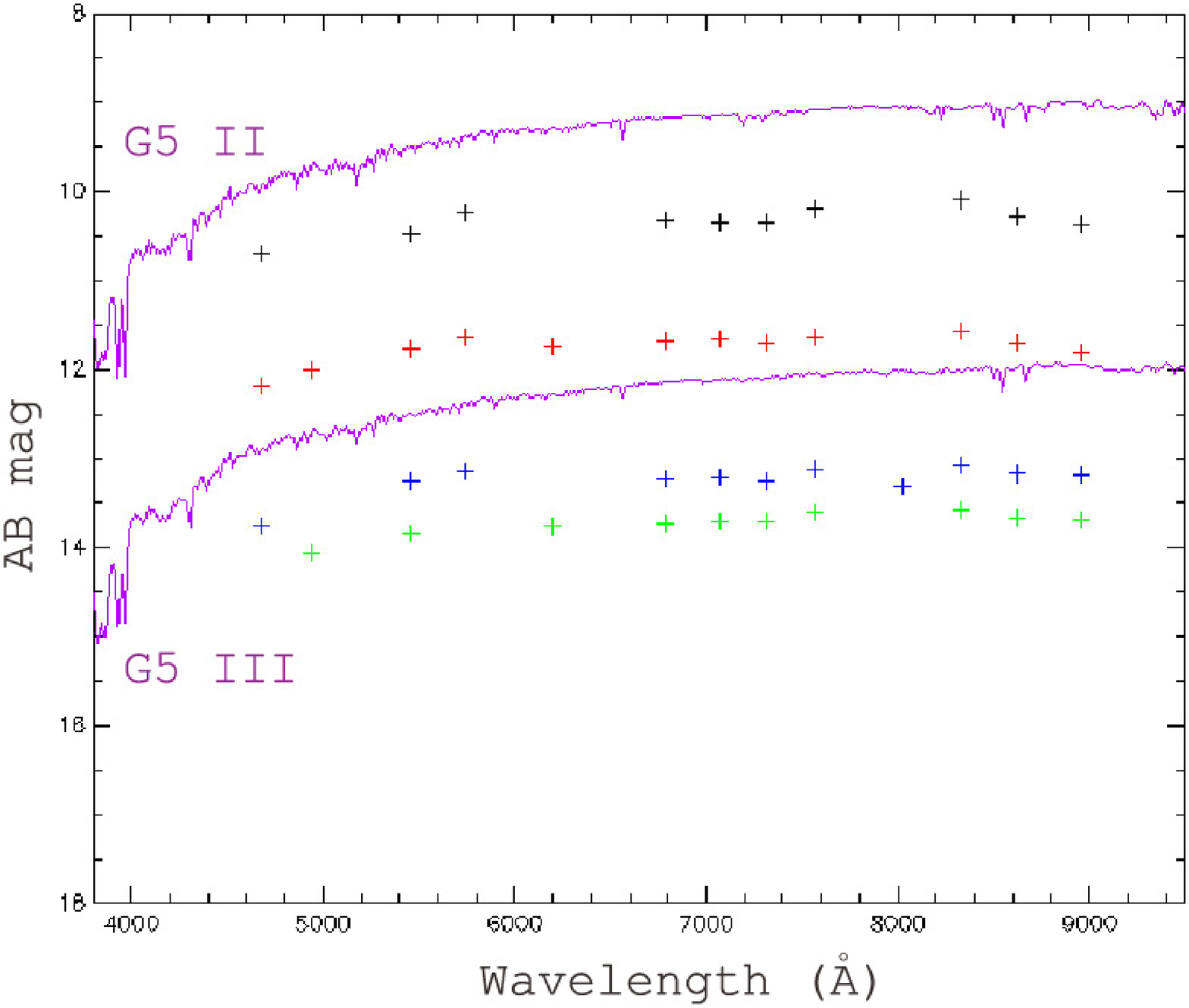}}}
 \caption[]{SED of brightest Ru 8 stars.}
 \label{sed8}
\end{figure}

\section{Summary}

We present the design of University of Tokyo's Dichoric-Mirror Camera (DMC). Two DMC observation runs were successfully carried out in 2007 and 2008, yielding scientific results. Although some results still need improvements, we are confident that DMC (and similar instruments later developed from this prototype) could obtain excellent results. Future possibilities include using Fabry-Perot mode to increase resolution, more observation of transient objects, and development of a more powerful version of the DMC.

\section*{Acknowledgment}
HK gratefully acknowledges the International Astronomical Union (IAU), Het Leids Kerkhoven-Bosscha-Fond (LKBF), and APRIM 2008 Organizing Committee for providing financial support which enabled him to participate in the conference. HK also wishes to thank the Tenmongaku Shinko Zaidan Fund for providing supports which enabled him to carry out the research in Japan. HLM would like to thank the Hitachi Scholarship Foundation for supporting his visit to Japan.

\label{lastpage}

\clearpage


\begin{thebibliography}{99}
\bibitem[\protect\citename{Bertelli et al. }1994]{bertelli94}Bertelli G., et al., 1994, A\&AS 106, 275
\bibitem[\protect\citename{Castellani et al. }2003]{castellani03}Castellani V., et al., 2003, A\&A 404, 645
\bibitem[\protect\citename{Doi et al. }1998]{doi98}Doi M., et al., 1998, Proc. SPIE 3355, 646
\bibitem[\protect\citename{Doi et al. }2008]{doi08}Doi M., et al., 2008, Proc. SPIE 7014, 70140F
\bibitem[\protect\citename{Drilling \& Landolt }2000]{drilling00}Drilling J.S. \& Landolt A.U., 2000, In Allen's Astrophysical Quantities, 4th ed., A.N. Cox, ed. Springer, Berlin
\bibitem[\protect\citename{Harris }1996]{harris96}Harris W.E., 1996, AJ 112, 1487
\bibitem[\protect\citename{Kuncarayakti }2008]{hk08}Kuncarayakti H., 2008, Master's thesis, Department of Astronomy, Institut Teknologi Bandung
\bibitem[\protect\citename{Pickles }1998]{pickles98}Pickles A.J., 1998, PASP 110, 863
\bibitem[\protect\citename{Rey et al. }2001]{rey01}Rey S.-C., et al., 2001, AJ 122, 3219
\bibitem[\protect\citename{Sekiguchi et al. }1992]{sekiguchi92}Sekiguchi M., et al., 1992, PASP 104, 744
\bibitem[\protect\citename{Utsunomiya }2008]{utsunomiya08}Utsunomiya H., 2008, Master's thesis, Department of Astronomy, Graduate School of Science, the University of Tokyo
\bibitem[\protect\citename{Wu et al. }2004]{wu04}Wu, Z.-Y., et al., 2004, PASP 117, 32
\end{thebibliography}
\end{document}